\documentclass[a4paper, 10pt, conference]{ieeeconf}      % Use this line for a4 paper

\IEEEoverridecommandlockouts                              % This command is only needed if
\usepackage{cite}
\usepackage{comment}
\usepackage{amsmath,amssymb,amsfonts}
\usepackage{algorithmic}
\usepackage[dvipdfmx]{graphicx}
\usepackage[dvipdfmx]{color}
\usepackage{graphicx}
\usepackage{color}
\usepackage{textcomp}
\usepackage[dvipdfmx]{xcolor}
\usepackage{url}
\usepackage{here}
\usepackage{listings}
\usepackage{arydshln}

\title{\LARGE \bf
Toward a Dialogue System Using a Large Language Model to Recognize User Emotions with a Camera
}

\author{Hiroki Tanioka$^{1\dag}$, Tetsushi Ueta$^{1}$ and Masahiko Sano$^{1}$% <-this % stops a space
\thanks{$^{1}$Center for Administration of Information Technology, Tokushima University, 770-8506 Tokushima, Japan
}%
\thanks{$^{\dag}${\tt\small tanioka.hiroki@tokushima-u.ac.jp}}%
}

\begin{document}

\maketitle
\thispagestyle{empty}
\pagestyle{empty}

%%%%%%%%%%%%%%%%%%%%%%%%%%%%%%%%%%%%%%%%%%%%%%%%%%%%%%%%%%%%%%%%%%%%%%%%%%%%%%%%
\begin{abstract}

The performance of ChatGPT\copyright{} and other LLMs has improved tremendously, and in online environments, they are increasingly likely to be used in a wide variety of situations, such as ChatBot on web pages, call center operations using voice interaction, and dialogue functions using agents.
In the offline environment, multimodal dialogue functions are also being realized, such as guidance by Artificial Intelligence agents (AI agents) using tablet terminals and dialogue systems in the form of LLMs mounted on robots.
In this multimodal dialogue, mutual emotion recognition between the AI and the user will become important. So far, there have been methods for expressing emotions on the part of the AI agent or for recognizing them using textual or voice information of the user's utterances, but methods for AI agents to recognize emotions from the user's facial expressions have not been studied.
In this study, we examined whether or not LLM-based AI agents can interact with users according to their emotional states by capturing the user in dialogue with a camera, recognizing emotions from facial expressions, and adding such emotion information to prompts. The results confirmed that AI agents can have conversations according to the emotional state for emotional states with relatively high scores, such as Happy and Angry.
\end{abstract}

%%%%%%%%%%%%%%%%%%%%%%%%%%%%%%%%%%%%%%%%%%%%%%%%%%%%%%%%%%%%%%%%%%%%%%%%%%%%%%%%
\section{Introduction}\label{sec:intro}

In August 2024, some teams of Center for Administration of Information Technology (abbreviated: AIT center) in Tokushima University will be moving out. Since the move will be to a different building, it was discussed to put up a sign or other information in the original room. Ideas for the signs included noting the location of the new rooms and a QR code for a Web page that would provide detailed directions. The idea is that it is effective to use both analog as well as digital methods.
However, we are the AIT center that should develop the most digital services within the university. The initial motivation for this study is to take this opportunity one step further and open an online contact point.

In order to set up an online window, digital devices (terminals) such as tablets are essential. Some suggested that a university character (Tokupon) could serve as a receptionist on standby.
If it goes that far, it would be like a receptionist robot. To realize this, a tablet terminal or similar device should be set up in front of the room before the move, ready for online meetings.
Online meetings can be conducted remotely by a representative waiting in the new room or by a pseudo-Artificial Intelligence agent (AI agent).
Therefore, we considered using ChatGPT\copyright{} one of the representatives of Large Language Models(LLMs), as the AI agent.

In this case, the person in charge of interacting remotely can see the other person's facial expression while conversing if the conversation is conducted in an online conference. However, LLMs usually recognize only textual information; some services, such as ChatGPT-4o~\cite{chatgpt-4o} and Gemini~\cite{gemini}, can recognize images and text simultaneously, but special prompts must be developed to capture them as the facial expressions of the interlocutor. In this study, we propose a method to recognize emotional information from the user's facial expression captured by a camera and hand it over to the LLM as a prompt together with the dialogue content.

\begin{figure}[ht]
\centerline{
\includegraphics[width=1.0\linewidth]{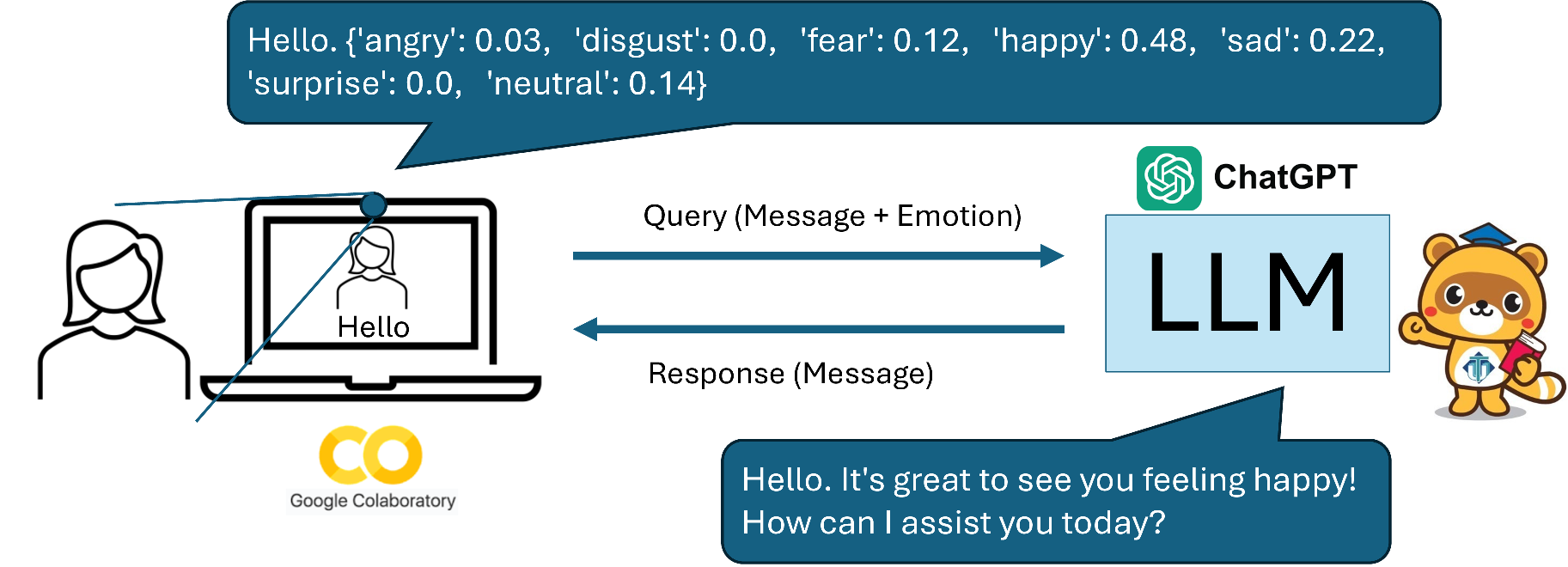}
%\hspace{0.1cm}
}
%\vspace{60mm}
%\caption{Comparison of non-significant Quiz 1.3 and significant problems Quiz 1.5}
\caption{Overview of FacingBot(FBot) system.}
\label{fig:approach}
%\vspace{3mm}
\end{figure}

\section{Related research}\label{sec:related}

\subsection{Emotion recognition technologies}
Recently, a technique called EmotionNet~\cite{7780969}, which applies convolutional neural networks to the recognition of emotions in human facial photos, has been proposed.
LibreFace~\cite{10484436} uses deep learning technology to recognize emotional expressions with higher accuracy than conventional methods.
In addition, techniques for recognizing emotions from images, such as those using the Vision Transformer, have been proposed by Fuyan Ma et al~\cite{9585378}.
Since ChatGPT-4o is a multimodal AI, it can be recognize emotions from facial expressions. However, in this study, we employed FER instead of ChatGPT-4o because of the need for image recognition technology that can work locally (or embedded), taking personal information into consideration.
FER is a Python library trained using Dataset~\cite{10.1007/978-3-642-42051-1_16}. The model is based on the Convolutional Neural Network proposed in Octavio Arriaga et al.~\cite{Arriaga2017RealtimeCN}.

\subsection{Multimodal interactive robot}

A survey article~\cite{10.1145/3648605}, which conducted a systematic review using the PRISMA protocol, found that task-based conversations are becoming more systematic, that there are improvements in the recognition rate of speech recognition, and that the appearance and expressiveness of the robot must be improved. The results of this study show that there is a need to improve the appearance and expressiveness of the robot. It is also reported that multimodal cues such as gestures, eye gaze, or facial expressions can be used.
Miyama et al.~\cite{doi:10.1080/01691864.2024.2319137} attempted to enrich task-based conversations with multimodal interaction by an android robot, with good results.
DFER-CLIP~\cite{7780969} attempts to use LLM to recognize emotions for multimodal information, while ChatGPT-4o~\cite{chatgpt-4o} and Gemini~\cite{gemini} are also being used to combine voice and image with text information. Multimodal conversations combining voice, images, and textual information are becoming possible.

\begin{figure*}[t]
\centerline{
\includegraphics[width=1.0\linewidth]{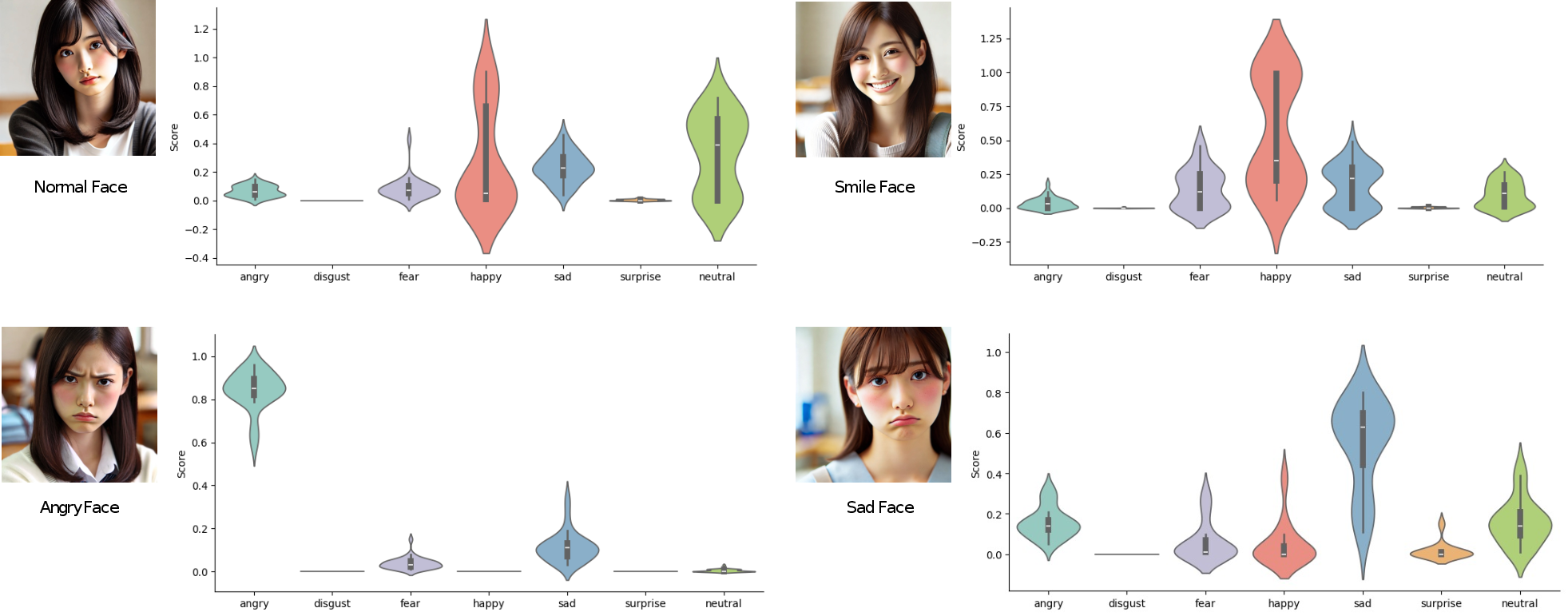}
}
\caption{Faces (upper left: normal, upper right: smiling, lower left: angry, and lower right: sad) of Japanese woman who visiting for consultation.}
\label{fig:faces}
%\vspace{3mm}
\end{figure*}

\section{Approach}\label{sec:approach}
Our proposed LLM-based facial expression recognition ChatBot is called FacingBot (FBot).
We employ a python library FER.
The FBot system configuration is shown in Figure~\ref{fig:approach}.
First, the user's face to be interacted with was generated by ChatGPT-4o using the face of a Japanese woman with four different facial expressions: normal, smiling, angry, and sad, as shown in Figure~\ref{fig:faces}.
The generated face images were printed out on paper and recognized in the form of a photograph taken with the built-in camera of a laptop computer.
Next, gpt-3.5-Turbo~\cite{gpt-3.5-turbo} was used as the LLM for dialogue in natural language. In addition to simply entering dialogue sentences, emotional information obtained using FER was added to the prompts in JSON(JavaScript Object Notation) format.

\subsection{Emotional expressions in JSON format}
Figure~\ref{fig:JSON} shows an example of recognition results for a single smiling image. ``angry'', ``disgust'', ``fear'', ``happy'', ``sad'', ``surprise'', and ``neutral'' each take a value between $0$ and $1$.
\begin{figure}[H]
\lstset{
    frame=single,
    %numbers=left,
    tabsize=2,
    basicstyle=\scriptsize
}
%\begin{verbatim}
\begin{lstlisting}[language=xml]
{
  'angry': 0.03,
  'disgust': 0.0,
  'fear': 0.12,
  'happy': 0.48,
  'sad': 0.22,
  'surprise': 0.0,
  'neutral': 0.14
}
\end{lstlisting}
%\end{verbatim}
\vspace{-2mm}
\caption{Example of emotional expressions for a smile face in JSON format.}
\label{fig:JSON}
\end{figure}

\subsection{Query prompt with JSON}
The prompt simply concatenated the message by string and the emotional information in JSON format, as expressed in the formula~\eqref{eq:prompt}.
\begin{equation}
 [\text{prompt}] = [\text{message}] + [\text{emotion(json)}] \label{eq:prompt}
\end{equation}
Figure~\ref{fig:prompt} is an example of a prompt when the message is ``Hello.'' and the query is made with a smile.
\begin{figure}[H]
\lstset{
    frame=single,
    %numbers=left,
    tabsize=2,
    basicstyle=\scriptsize
}
%\begin{verbatim}
\begin{lstlisting}[language=xml]
'role': 'user',
'content': 'Hello.' + '('
{
  'angry': 0.03,
  'disgust': 0.0,
  'fear': 0.12,
  'happy': 0.48,
  'sad': 0.22,
  'surprise': 0.0,
  'neutral': 0.14
} + ')'
\end{lstlisting}
%\end{verbatim}
\vspace{-2mm}
\caption{Example of user prompt with JSON (for a smile face).}
\label{fig:prompt}
\end{figure}
\begin{figure}[H]
\centerline{
\includegraphics[width=0.8\linewidth]{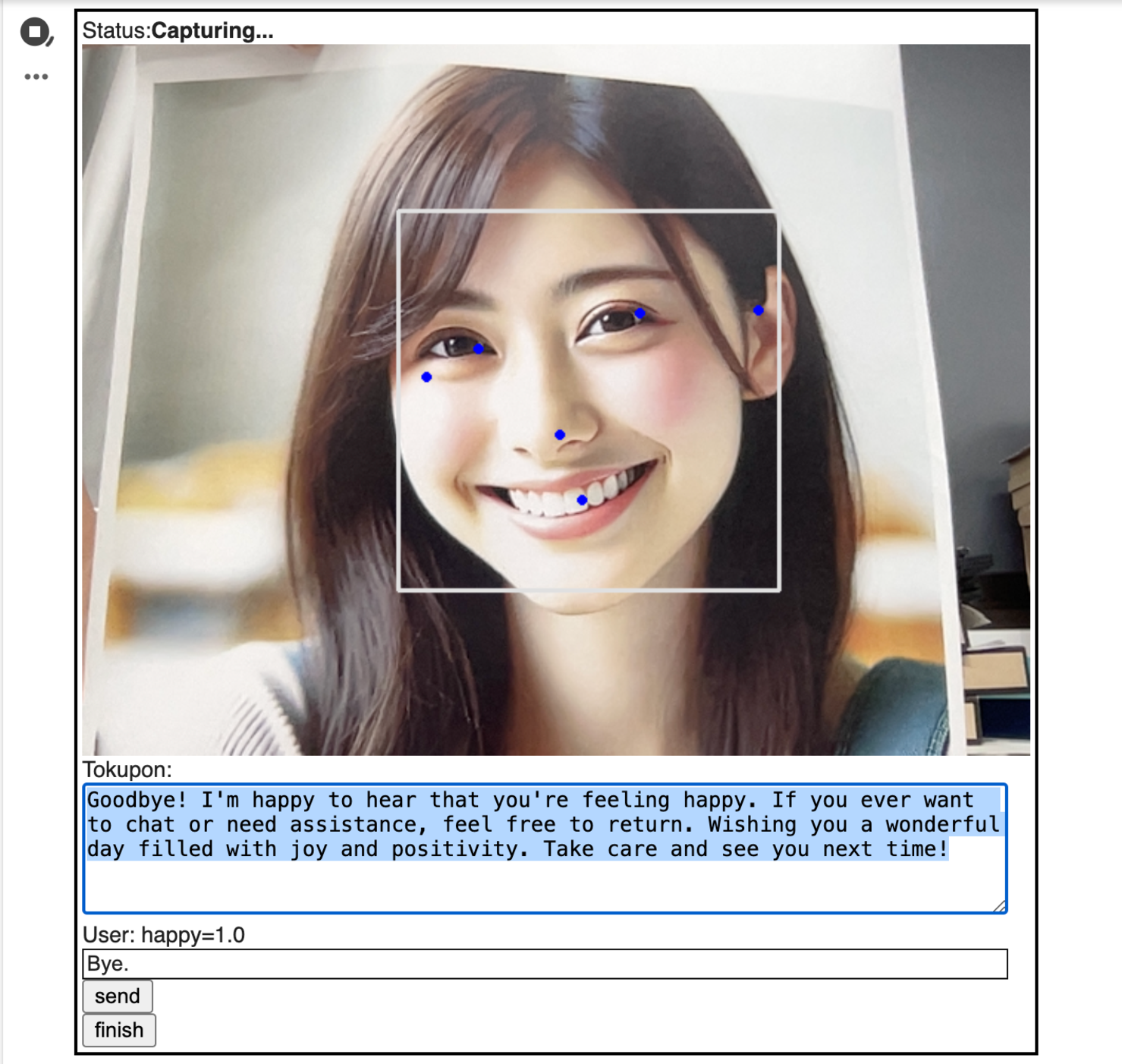}
%\hspace{0.1cm}
}
%\vspace{60mm}
%\caption{Comparison of non-significant Quiz 1.3 and significant problems Quiz 1.5}
\caption{Screenshot of the proposed system execution screen}
\label{fig:happy_bye}
%\vspace{-1mm}
\end{figure}

\section{Experiment}\label{sec:experiment}

First, prepare multiple face images of one user interacting with gpt-3.5-turbo. Next, prepare several scenarios of interaction with gpt-3.5-turbo.
Then, we will import and implement the FER and OpenAI libraries on Google Colaboratory using Python 3.10.12.
In the experiment, the user interacts with gpt-3.5-turbo according to the scenarios, and we will investigate how different facial expressions affect gpt-3.5-turbo's response.

\subsection{Preparation of face images}\label{sec:images}

Using ChatGPT-4o, four facial expressions were generated for a Japanese woman who came for a consultation. In this case, four types of facial expressions are used: Normal, which has no emotional expression, a smiling face, an angry face, and a sad face, simply because she is asking for advice.
They were captured using the built-in camera of a laptop computer, and the results of emotion recognition by FER are shown in Figure~\ref{fig:faces}.
From the images, we found that women's facial expressions are either Happy or Neutral when normal, but almost Happy when smiling, Angry when angry, and Sad when sad, with higher score values for sad faces.

\subsection{Dialogue scenario}\label{sec:dialogue}

Two dialogue scenarios were used: Case A was simply saying ``Hello.'' and by looking at the differences in FBot's responses when spoken to with a normal expression, a smile, an angry face, or a sad face, we can see if the emotional expressions added in JSON format were correctly recognized.
In the other case B, the user first asks ``How can I comfort a friend with a broken heart?'' and responds to FBot's answer with ``Thank you.'' followed by ``Bye.''
At this point, we see what kind of response we get for the four different facial expressions.

\begin{itemize}
  \item Case A
  \begin{description}
    \item[User] ``Hello.'' (A1)
    \item[FBot] (Answer for A1)
  \end{description}
  \item Case B
  \begin{description}
    \item[User] ``How can I comfort a friend with a broken heart?'' (B1)
    \item[FBot] (Answer for B1)
    \item[User] ``Thank you.'' (B2)
    \item[FBot] (Answer for B2)
    \item[User] ``Bye.'' (B3)
    \item[FBot] (Answer for B3)
  \end{description}
\end{itemize}

\subsection{Results}\label{sec:result}

Figure~\ref{fig:happy_bye} shows how the user says ``Bye.'' with a smile in response to FBot.
As shown in Table~\ref{tab:results}, when spoken to with a smile, the response is that the speaker is pleased to be smiling.
For instance, if the user inputed ``Hello.'' with smile face, FBot responded ``I'm glad you are happy! How can I help you today?''
On the other hand, if the face is angry or sad, there is a caring remark.
For instance, if the user inputed ``Hello.'' with angry face, FBot responded ``I'm here to listen and support you. If you'd like to talk about what's on your mind or if there's anything I can do to help, feel free to share. Remember, it's okay to feel a range of emotions, and I'm here to offer assistance in any way I can. Take a deep breath, and know that you're not alone.''
These results confirm that it is possible to create a dialogue based on the user's emotional state by adding the emotion information recognized by FER in JSON format to gpt-3.5-turbo and sending it to a conversation using text strings.

\begin{table*}[htbp]
%\small
%\scriptsize
\tiny
\caption{Results of Dialogue Experiments with Emotional Information}
\label{tab:results}
\begin{center}
\begin{tabular*}{17.5cm}{ccp{7em}cp{49em}ccc}
\hline \hline
Case & Number & Query & Face & Response (beginning of example) & Understanding & Worrying & Encouraging \\
\hline \hline
A & 1 & Hello.& Normal &
Hello! It's nice to hear from you. How are you feeling today? If there's something on your mind or if you'd like to chat about anything, I'm here to listen and support you. Remember, your emotions are valid, and it's okay to express how you feel. Take your time, and feel free to share whatever you're comfortable with.
& - & - & - \\ \cdashline{5-8}
 &  &  & Smile & Hello. It's great to see you \textbf{feeling happy}! How can I assist you today? & \checkmark & - & - \\ \\ \cdashline{5-8}
 &  &  & Angry &
Hello! I'm here to listen and support you. If you'd like to talk about what's on your mind or if there's anything I can do to help, feel free to share. Remember, it's okay to feel a range of emotions, and I'm here to offer assistance in any way I can. \textbf{Take a deep breath, and know that you're not alone.}
& \checkmark & \checkmark & \checkmark \\ \cdashline{5-8}
 &  &  & Sad &
Hello! I'm here for you. If you'd like to talk about what's been bothering you or if there's anything specific you need help with, please feel free to share. Remember, it's important to express your emotions and seek support when \textbf{you're feeling down.} I'm here to listen and offer my assistance in any way I can. \textbf{Take your time, and know that you are not alone.}
& \checkmark & \checkmark & \checkmark \\
\hline
B & 1 & How can I comfort a friend with a broken heart? & Normal & I'm sorry to hear that your friend is going through a tough time. When comforting a friend with a broken heart, it's important to listen to them without judgment and offer your support. You can let them know that you're there for them, and remind them that it's okay to feel the way they do. Encouraging them to express their emotions and providing a shoulder to lean on can help them feel less alone in their pain. Just being there for your friend and showing that you care can make a big difference. & - & - & - \\
\cline{2-8}
 & 2 & Thank you. & Normal &
You're welcome! I'm here to support you. If you ever need to talk more about comforting your friend or if there's anything else on your mind, feel free to reach out. Remember, your compassion and care for your friend are truly valuable. \textbf{Take care of yourself too during this time.} Sending you positive thoughts and warm wishes.
& - & - & \checkmark \\ \cdashline{5-8}
 & 3 & Bye. & Normal &
Goodbye! If you ever feel like talking or need a listening ear, don't hesitate to come back. Remember to take care of yourself and know that brighter days are ahead. \textbf{Stay strong and keep a positive mindset.} Sending you virtual support and best wishes. \textbf{Take care!}
& - & \checkmark & \checkmark \\ \cline{2-8}
 & 2 & Thank you. & Smile &
You're welcome! I'm glad to see you \textbf{feeling happy}. If you have any more questions or need further assistance in the future, feel free to reach out. Remember, your kindness and support can truly brighten someone's day. \textbf{Keep spreading positivity and joy. Take care and stay happy!}
 & \checkmark & - & \checkmark \\ \cdashline{5-8}
 & 3 & Bye. & Smile &
Goodbye! I'm happy to hear that \textbf{you're feeling happy}. If you ever want to chat or need assistance, feel free to return. Wishing you a wonderful day filled with joy and positivity. \textbf{Take care and see you next time!}
 & \checkmark & - & \checkmark \\ \cline{2-8}
 & 2 & Thank you. & Angry &
You're welcome! It's important to acknowledge your feelings and know that it's okay to experience a range of emotions. If you ever need someone to talk to or seek comfort, don't hesitate to reach out. Remember, \textbf{you are not alone, and there are people who care about you and want to support you.} \textbf{Take care of yourself and remember to be kind to yourself during this time.} \textbf{Sending you positive thoughts and strength.}
& \checkmark & \checkmark & \checkmark \\ \cdashline{5-8}
 & 3 & Bye. & Angry &
Goodbye! If you ever feel like talking or need a listening ear, don't hesitate to come back. \textbf{Remember to take care of yourself and know that brighter days are ahead.} \textbf{Stay strong and keep your head up.} Sending you positive vibes and warm wishes. \textbf{Take care!}
 & - & \checkmark & \checkmark \\ \cline{2-8}
 & 2 & Thank you. & Sad &
You're welcome! I'm here to support you and offer guidance whenever you need it. Remember that \textbf{you are a caring friend, and your efforts to comfort and support others are truly admirable.} If you ever want to talk more or share your feelings, feel free to reach out. \textbf{Keep spreading kindness and positivity, even during challenging times.} \textbf{Take care, and stay strong.} Sending you virtual hugs and support!
& \checkmark & \checkmark & \checkmark \\ \cdashline{5-8}
 & 3 & Bye. & Sad &
Goodbye! If you ever feel like talking or need a listening ear, don't hesitate to come back. \textbf{Take care of yourself and remember that brighter days are ahead.} Sending you positive vibes and warm wishes. See you next time!
 & - & - & \checkmark \\
\hline \hline
\multicolumn{5}{l}{Hyphen (-) means Unknown.}
\end{tabular*}
\end{center}
\end{table*}

\section{Discussion}\label{sec:discussion}
The following is a discussion of the experimental results. First, we confirmed that gpt-3.5-turbo can recognize emotional states from users' facial expressions using FER.
Next, we confirmed that gpt-3.5-turbo can respond to multimodal queries with emotion information according to the emotional state.
We also confirmed that gpt-3.5-turbo does not respond according to the emotional state when the emotional state is ambiguous.

Other issues that were found through the experiments are that emotion recognition using FER fluctuates the recognition results depending on the proximity, brightness, and angle of the face. Furthermore, there are still issues to be considered, such as whether to acquire the average or the maximum value of the emotion information within a certain period of time before the chat program speaks to the user. The average facial expression within a certain time period needs to be calculated, or the facial expression immediately after the chat string is sent may be important. This area needs to be investigated and studied, and the system design needs to be reviewed.

\section{Conclusion}\label{sec:conclusion}
In this paper, for multimodal interaction by AI agents, we designed and developed a system that, when a user makes an inquiry offline, sends visual information, especially emotional information obtained from facial expressions, along with the inquiry to the LLM to ensure that the system can respond according to the facial expressions of the user.
%.
Experimental results showed that LLM can respond differently to smiling, angry, and sad faces respectively.
However, further research is needed on how to summarize the emotional information added to the text of the query by the user, since the facial expressions of the user are not always constant and the recognition results by FER are also variable.
Based on the results of this research, the system could be applied to offline receptionist operations using AI agents, online receptionist operations using AI agents, information sharing systems in university laboratories, and university operations such as medical institutions.
It could also be used to understand the psychological state of athletes and coaches from their facial expressions, to understand the state of athletes and teams, and to improve the training efficiency of athletes.
In the field of information security, it may be possible to determine whether a person is suspicious or not from his/her facial expression by combining visual information and query information in reception work that requires identification, such as password re-issuance at universities and companies.

%.
In the future, we would like to develop an algorithm to more accurately grasp emotions from facial expressions, develop a learning model to recognize the emotions of users wearing masks, and conduct research and development of a multimodal dialogue system that combines speech recognition.
%\section*{APPENDIX}
%
%Appendixes should appear before the acknowledgment.

\section*{Acknowledgment}
We would like to thank our colleagues for cooperating to collect dataset and discussion. This work was supported by JSPS KAKENHI Grant Number JP22K12293 and JP18K11572.
Additionally, DeepL\copyright{} was utilized for translation purposes, ensuring the accuracy and quality of English content. %I am also grateful to my colleagues who helped me with my experiments in between work duties and to my family for giving me time to write my dissertation without getting emotional.

%%%%%%%%%%%%%%%%%%%%%%%%%%%%%%%%%%%%%%%%%%%%%%%%%%%%%%%%%%%%%%%%%%%%%%%%%%%%%%%%

\bibliographystyle{IEEEtran}
\bibliography{IEEEabrv, tanioka}

\end{document}